\documentclass[12pt]{article}
 \usepackage[cp1251]{inputenc} 
 \usepackage[english, russian]{babel} 
 \usepackage{amsmath, latexsym, amssymb, bm, array, graphics, eucal,
 amsfonts,}
 \makeatletter
\newcommand{\const}{\mathop{\rm const\, }}
\renewcommand{\Re}{\mathop{\rm Re}}
\renewcommand{\Im}{\mathop{\rm Im\,}}

 \makeatother
 
\usepackage[dvips]{graphicx}

\begin{document}
\thispagestyle{empty}
\large
\renewcommand{\abstractname}{Abstract}
\renewcommand{\refname}{\begin{center}
 REFERENCES\end{center}}
\newcommand{\mc}[1]{\mathcal{#1}}
\newcommand{\E}{\mc{E}}
\makeatother

\begin{center}
\bf Theory of orthogonality of eigenfunctions  of the
characteristic equations as a method of solution boundary
problems for model kinetic equations
\end{center} \medskip

\begin{center}
  \bf
  A. V. Latyshev\footnote{$avlatyshev@mail.ru$},
  A. D. Kurilov\footnote{$ad.kurilov@gmail.com$}
\end{center}\medskip

\begin{center}
{\it Faculty of Physics and Mathematics,\\ Moscow State Regional
University, 105005,\\ Moscow, Radio str., 10A}
\end{center}\medskip

\begin{abstract}
We consider two classes of linear
kinetic equations: with constant collision frequency and
constant mean free path of gas molecules (i.e.,
frequency of molecular collisions, proportional to the modulus
molecular velocity). Based  homogeneous Riemann boundary value problem
with a coefficient equal to the ratio of the boundary values
dispersion function, develops the theory of the half-space
orthogonality of generalized singular eigenfunctions
corresponding characteristic equations, which leads
separation of variables.

And in this two boundary value problems of the kinetic theory (diffusion
light component of a binary gas and Kramers problem about
isothermal slip) shows the application of the theory
orthogonality eigenfunctions for analytical solutions
these tasks.

{\bf Key words:} kinetic equation,
collision frequency, boundary value problems, eigenfunctions,
dispersion function, analytical solution.

\medskip

PACS numbers:  05.60.-k   Transport processes,
51.10.+y   Kinetic and transport theory of gases,

\end{abstract}

\begin{center}
\bf 1.  Introduction
\end{center}

Construction of precise solutions of boundary value problems
mathe\-ma\-ti\-cal physics is a great success. This fully
applies to the boundary value problems for kinetic equations.

In 1960 K. Case in his work \cite{1} for the first time proposed a
method of analytical solutions of boundary value problems for
the model equation neutron transport
$$
\mu\dfrac{\partial h}{\partial x}+h(x,\mu)=\dfrac{c}{2}
\int\limits_{-1}^{1}h(x,\mu')d\mu'.
\eqno{(1.1)}
$$

The general method of Fourier's separation of variables leads
to the substitution
$$
h_\eta(x,\mu)=\exp\Big(-\dfrac{x}{\eta}\Big)\Phi(\eta,\mu).
\eqno{(1.2)}
$$
Substituting (1.2) reduces equation (1.1) to the characteristic
equation
$$
(\eta-\mu)\Phi(\eta,\mu)=\eta\dfrac{c}{2}
\int\limits_{-1}^{1}\Phi(\eta,\mu')d\mu'.
\eqno{(1.3)}
$$

K. Case's brilliant hunch was that it offered seek a solution
of the characteristic equation (1.3) in space of generalized
functions \cite{Vladimirov}
$$
\Phi(\eta,\mu)=\eta\dfrac{c}{\sqrt{\pi}}P\dfrac{1}{\eta-\mu}+
\lambda(\eta)\delta(\eta-\mu),
\eqno{(1.4)}
$$
where $\lambda(z)$ is the dispersion function,
$$
\lambda(z)=1+\dfrac{z}{2}\int\limits_{-1}^{1}\dfrac{d\tau}{\tau-z},
$$
$P x^{-1}$ is the generalized function (principal value of the integral
in the integration $x^{-1}$), $\delta(x)$ is the Dirac delta function.

Properties of the eigenfunctions (1.4), expansion of the
solutions of equations (1.1) and their generalizations in
eigenfunctions were inves\-ti\-ga\-ted in works \cite{2}--\cite{8}.

One of the first boundary value problems for a model kinetic
BGK equation (Bhatnagar, Gross, Krook), for which was
exact solution is obtained, has been linearized problem of the Kramers
isothermal slip. This problem was solved analytically in 1962 C.
Cercignani \cite{9}.

After  Cercignani's work have been numerous attempts to solve
analy\-ti\-cally the Smoluchowski problem of the temperature jump
and low evaporation. An overview of such attempts is presented
in the works \cite{10}--\cite{12}. These attempts continued until
the appearance of work \cite{13}, which was developed analytical
method of solving boundary value problems for this class of kinetic
equations, which can be reduced to the solution vector
integro-differential equations of the type transport equations.

С. Cercignani \cite{9} reduced the solution of the isothermal
slip problem to solving the following boundary value problem
$$
\mu\dfrac{\partial h}{\partial x}+h(x,\mu)=\dfrac{1}{\sqrt{\pi}}
\int\limits_{-\infty}^{\infty}e^{-\mu'^2}h(x,\mu')d\mu',\quad
x>0,-\infty<\mu<+\infty,
$$
$$
h(0,\mu)=0,\qquad \mu>0,
$$
$$
h(x,\mu)=h_{as}(x,\mu)+o(1), \qquad x\to +\infty.
$$

Here $h_{as}(x,\mu)$ is the Chapman---Enskog asymptotic distribution
func\-tion,
$$
h_{as}(x,\mu)=2U_0+2G_v(x-\mu),
$$
$U_0$ is the unknown dimensionless slip velocity gas, subject to
finding, $G_v$ is the specified far from the wall dimensionless
mass velocity gradient, $\mu=C_x$, ${\bf C}={\bf v}/v_T$, $v_T=1/
\sqrt{\beta}$ is the thermal velocity of the gas, $\beta=m/(2kT)$,
$m$ is the mass of gas molecule, $k$ is the Boltzmann constant,
$T=\const$ is the gas temperature.

In the problem of evaporation of the binary gas light component (see,
e.g., \cite{14}) investigated the one-parameter family of equations
$$
\mu\dfrac{\partial h}{\partial x}+h(x,\mu)=\dfrac{c}{\sqrt{\pi}}
\int\limits_{-\infty}^{\infty}e^{-\mu'^2}h(x,\mu')d\mu',
\eqno{(1.5)}
$$
where $c$ is the numeric parameter, $0<c<1$,
$x>0$,$-\infty<\mu<+\infty$.

In works \cite{15} and \cite{16} in solving boundary value
problems for a model kinetic equation with the collision
frequency, proportional to the modulus of molecular speed,
consider the equation
$$
\mu\dfrac{\partial h}{\partial x}+h(x,\mu)=\dfrac{3}{4}
\int\limits_{-1}^{1}(1-\mu'^2)h(x,\mu')d\mu',\quad
x>0,-1<\mu<+1.
\eqno{(1.6)}
$$

In this paper we develop the theory of orthogonality
eigenfunctions of the characteristic equations corresponding
equations (1.5) and (1.6). Underlying this theory is the
solution of the boundary Riemann problem \cite{Gakhov}
from the theory of complex variable functions.
This theory is then applied to the solution of boundary
value problems for equations (1.5) and (1.6).

\begin{center}
{\bf 2. Eigenfunctions in the problem of the diffusion of
the binary gas light component and their orthogonality}
\end{center}

We consider the equation (1.5). The general method of Fourier's
separation of variables, as already mentioned, leads to the
substitution
$$
h_\eta(x,\mu)=e^{-x/\eta}\Phi(\eta,\mu),
\eqno{(2.1)}
$$
where $\eta$ is the complex-valued spectral parameter.

Substituting (2.1) in (1.5), immediately obtain the
characteristic equa\-tion
$$
(\eta-\mu)\Phi(\eta,\mu)=\eta\dfrac{c}{\sqrt{\pi}}
\int\limits_{-\infty}^{\infty}e^{-\mu'^2}\Phi(\eta,\mu')d\mu'.
\eqno{(2.2)}
$$

We denote
$$
n(\eta)=\int\limits_{-\infty}^{\infty}e^{-\mu'^2}\Phi(\eta,\mu')d\mu'
\eqno{(2.3)}
$$
and rewrite (2.2) in the form
$$
(\eta-\mu)\Phi(\eta,\mu)=\eta\dfrac{c}{\sqrt{\pi}}n(\eta).
\eqno{(2.4)}
$$

By the homogeneity of the equation (1.5) without loss of
generality, we can assume that
$$
n(\eta)\equiv\int\limits_{-\infty}^{\infty}e^{-\mu'^2}
\Phi(\eta,\mu')d\mu'=1.
\eqno{(2.5)}
$$

From equations (2.3) and (2.5) in the space of generalized
functions
\cite{Vladimirov} we find the eigenfunctions corresponding to
the continuous spectrum
$$
\Phi(\eta,\mu)=\eta\dfrac{c}{\sqrt{\pi}}P\dfrac{1}{\eta-\mu}+
e^{\eta^2}\lambda_c(\eta)\delta(\eta-\mu).
\eqno{(2.6)}
$$

Where $\lambda_c(\eta)$ is the dispersion function,
$$
\lambda_c(\eta)=1+z\dfrac{c}{\sqrt{\pi}}\int\limits_{-\infty}^{\infty}
\dfrac{e^{-\tau^2}d\tau}{\tau-z},
$$

The basic theory of orthogonality we set scalar product with
weight $\rho(\mu)=e^{-\mu^2}\gamma(\mu)$, where
$$
\gamma(\mu)=\mu\dfrac{X^+(\mu)}{\lambda_c^+(\mu)}.
$$

Here $X(z)$ is the solution of the homogeneous Riemann boundary
value problem from \cite{14}
$$
\dfrac{X^+(\mu)}{X^-(\mu)}=\dfrac{\lambda^+_c(\mu)}{\lambda_c^-(\mu)},
\quad \mu>0.
$$
The solution of this problem (see \cite{14}) defined by the equalities
$$
X(z)=\dfrac{1}{z}e^{V(z)}, \qquad
V(z)=\dfrac{1}{\pi}\int\limits_{0}^{\infty}\dfrac{\theta(\mu)-\pi}
{\mu-z}d\mu,
$$
$$
\theta(\mu)=\arg\lambda^+_c(\mu)=\arcctg\dfrac{\Re \lambda_c^+(\mu)}
{\Im\lambda_c^+(\mu)},\qquad \theta(0)=0,
$$
$$
\lambda_c^+(\mu)=\Re\lambda_c^+(\mu)+i\Im \lambda_c^+(\mu)=
\lambda_c(\mu)+ic\sqrt{\pi}\mu e^{-\mu^2},
$$
$$
\lambda_c(\mu)=1-2c
\mu^2e^{-\mu^2}\int\limits_{0}^{1}e^{\mu^2\tau^2}d\tau.
$$

Scalar product on the set of functions, that depend on the
speed variable $\mu$, we introduce by equality
$$
(f,g)=\int\limits_{0}^{\infty}e^{-\mu^2}\gamma(\mu)f(\mu)g(\mu)d\mu.
$$

For convenience the eigenfunctions $\Phi(\eta,\mu)$
we denote by $\Phi_\eta(\mu)$.

{\sc Theorem 1.}  Scalar product number one and eigenfunction
of the continuous spectrum is equal to the spectral parameter, i.e.
$$
(1,\Phi_\eta)=\eta,\qquad \eta>0.
\eqno{(2.8)}
$$

{\sc Proof.} By the definition of scalar product, we have
$$
(1,\Phi_\eta)=\int\limits_{0}^{\infty}e^{-\tau^2}
\gamma(\tau)\Phi_\eta(\tau)d\tau.
$$

We represent this expression in explicit form
$$
(1,\Phi_\eta)=\int\limits_{0}^{\infty}e^{-\tau^2}
\gamma(\tau)\Big[\dfrac{c\eta}{\sqrt{\pi}}P\dfrac{1}{\eta-\tau}+
e^{\eta^2}\lambda_c(\eta)\delta(\eta-\tau)\Big]d\tau=
$$
$$
=-\dfrac{c\eta}{\sqrt{\pi}}\int\limits_{0}^{\infty}
\dfrac{e^{-\tau^2}\gamma(\tau)d\tau}{\tau-\eta}+
\gamma(\eta)\lambda_c(\eta)\theta_+(\eta),
$$
where $\theta_+(\eta)$ is the Heaviside step function.

Now we use the integral representation (see \cite{14})
$$
X(z)=1+\dfrac{c}{\sqrt{\pi}}\int\limits_{0}^{\infty}
\dfrac{e^{-\tau^2}\gamma(\tau)}{\tau-z}d\tau.
$$

Using this representation, we obtain
$$
(1,\Phi_\eta)=-\eta X(\eta)+\eta+\eta\dfrac{X^+(\eta)}{\lambda_c(\eta)}
\dfrac{\lambda_c^+(\eta)+\lambda_c^-(\eta)}{2}=\eta,
$$
Q.E.D.

{\sc Theorem 2.} Eigenfunctions $\Phi_\eta(\mu)$ form an
orthogonal family and we have the equality
$$
(\Phi_\eta, \Phi_{\eta'})=N(\eta)\delta(\eta-\mu),
\eqno{(2.7)}
$$
where
$$
N(\eta)=e^{\eta^2}\gamma(\eta)\lambda_c^+(\eta)\lambda_c^-(\eta).
\eqno{(2.8)}
$$

{\sc Proof.} By the definition of scalar product, we have
$$
(\Phi_\eta, \Phi_{\eta'})=\int\limits_{0}^{\infty}e^{-\tau^2}
\gamma(\tau)\Phi_\eta(\tau)\Phi_{\eta'}(\tau)d\tau.
$$
We represent this expression in explicit form
$$
(\Phi_\eta, \Phi_{\eta'})=\int\limits_{0}^{\infty}e^{-\tau^2}
\gamma(\tau)\Big[\dfrac{c\eta}{\sqrt{\pi}}P\dfrac{1}{\eta-\tau}+
e^{\eta^2}\lambda_c(\eta)\delta(\eta-\tau)\Big]\times
$$
$$
\times\Big[\dfrac{c\eta'}{\sqrt{\pi}}P\dfrac{1}{\eta'-\tau}+
e^{\eta'^2}\lambda_c(\eta')\delta(\eta'-\tau)\Big]d\tau=
J_1+J_2+J_3+J_4.
$$

Here
$$
J_1=c^2\dfrac{\eta\eta'}{\pi}\int\limits_{0}^{\infty}
\dfrac{e^{-\tau^2}\gamma(\tau)d\tau}{(\eta-\tau)(\eta'-\tau)},
$$
$$
J_2=\dfrac{c\eta}{\sqrt{\pi}}e^{\eta'^2}\lambda_c(\eta')
\int\limits_{0}^{\infty}
\dfrac{e^{-\tau^2}\gamma(\tau)\delta(\eta'-\tau)}{\eta-\tau}d\tau,
$$
$$
J_2=\dfrac{c\eta'}{\sqrt{\pi}}e^{\eta^2}\lambda_c(\eta)
\int\limits_{0}^{\infty}
\dfrac{e^{-\tau^2}\gamma(\tau)\delta(\eta-\tau)}{\eta'-\tau}d\tau,
$$
$$
J_4=e^{\eta^2+\eta'^2}\lambda_c(\eta)\lambda_c(\eta')
\int\limits_{0}^{\infty}
e^{-\tau^2}\gamma(\tau)\delta(\eta-\tau)\delta(\eta'-\tau)d\tau.
$$

Second, third and fourth integrals are easily calculated as
convolution with the Dirac delta function
$$
J_2=\dfrac{c\eta\lambda_c(\eta')\gamma(\eta')}{\sqrt{\pi}(\eta-\eta')},
$$
$$
J_3=\dfrac{c\eta'\lambda_c(\eta)\gamma(\eta)}{\sqrt{\pi}(\eta'-\eta)},
$$
$$
J_4=e^{\eta^2}\lambda_c^2(\eta)\gamma(\eta)\delta(\eta-\eta').
$$

Calculate the first integral. We use the expansion into
elementary fractions
$$
\dfrac{1}{(\eta-\tau)(\eta'-\tau)}=\dfrac{1}{\eta-\eta'}
\Big(\dfrac{1}{\tau-\eta}-\dfrac{1}{\tau-\eta'}\Big),
$$
and the Poincar\' e---Bertrand formula
$$
P\dfrac{1}{\eta-\mu}P\dfrac{1}{\eta'-\mu}=P\dfrac{1}{\eta-\eta'}
\Big(P\dfrac{1}{\eta'-\mu}-P\dfrac{1}{\eta-\mu}\Big)+
$$
$$
+\pi^2 \delta(\eta-\mu)\delta(\eta'-\mu).
$$
As a result, we obtain
$$
J_1=c^2\dfrac{\eta\eta'}{\sqrt{\pi}}\Bigg[\dfrac{1}{\eta-\eta'}
\int\limits_{0}^{\infty}\dfrac{e^{-\tau^2}\gamma(\tau)d\tau}
{\tau-\eta}-\dfrac{1}{\eta-\eta'}
\int\limits_{0}^{\infty}\dfrac{e^{-\tau^2}\gamma(\tau)d\tau}
{\tau-\eta'}+
$$
$$
+\pi^2\int\limits_{0}^{\infty}e^{-\tau^2}\gamma(\tau)\delta(\eta-\tau)
\delta(\eta'-\tau)d\tau\Bigg].
$$
Now we use the integral representation \cite{14}
$$
X(z)=1+\dfrac{c}{\sqrt{\pi}}\int\limits_{0}^{\infty}
\dfrac{e^{-\tau^2}\gamma(\tau)}{\tau-z}d\tau.
$$
With its help the integral $J_1$ is equal to
$$
J_1=c\dfrac{\eta\eta'}{\sqrt{\pi}}\dfrac{X(\eta)-X(\eta')}{\eta-\eta'}+
\pi c^2\eta^2e^{-\eta^2}\gamma(\eta)\delta(\eta-\eta').
$$

We find the sum
$$
J_1+J_2=\dfrac{c}{\sqrt{\pi}}\dfrac{\eta\lambda_c(\eta')\gamma(\eta')-
\eta'\lambda_c(\eta)\gamma(\eta)}{\eta-\eta'}.
$$

We use the definition of the function $\gamma(\tau)$. Then we see that
$$
J_2+J_3=\dfrac{c\eta\eta'}{\sqrt{\pi}(\eta-\eta')}\Big[\lambda_c(\eta')
\dfrac{X^+(\eta')}{\lambda_c^+(\eta')}-\lambda_c(\eta)
\dfrac{X^+(\eta)}{\lambda_c^-(\eta)}\Big].
$$

Value of the dispersion function at the cut in this equality we
will replace half the sum of its boundary values, as well as we use
homoge\-ne\-ous Riemann boundary value problem. The result is that
$$
J_2+J_3=\dfrac{c\eta\eta'}{\sqrt{\pi}(\eta-\eta')}\Big[
\dfrac{X^+(\eta')+X^-(\eta')}{2}-\dfrac{X^+(\eta)+X^-(\eta)}{2}\Big]=
$$
$$
=\dfrac{c\eta\eta'}{\sqrt{\pi}(\eta-\eta')}[X(\eta')-X(\eta)].
$$

Adding expressions $J_1$, $J_2+J_3$ and $J_4$, we see that
$$
(\Phi_\eta,\Phi_{\eta'})=\Big[e^{\eta^2}\lambda_c^2(\eta)+
c^2\pi \eta^2e^{-\eta^2}\Big]\gamma(\eta)\delta(\eta-\eta')=
$$
$$
=[\lambda_c(\eta)+i\sqrt{\pi}c\eta e^{-\eta^2}]
[\lambda_c(\eta)-i\sqrt{\pi}c\eta e^{-\eta^2}]e^{\eta^2}\gamma(\eta)
\delta(\eta-\eta')=
$$
$$
=\lambda_c^+(\eta)\lambda_c^-(\eta)e^{\eta^2}\gamma(\eta)
\delta(\eta-\eta')=|\lambda_c^+(\eta)|^2e^{\eta^2}\gamma(\eta)
\delta(\eta-\eta')=
$$
$$
=N(\eta)\delta(\eta-\eta'),
$$
Q.E.D.

We apply the theorem to solve the problem of the diffusion
of the binary gas light component. In \cite{14} shows
that the solution of this problem reduces to the solution
of the integral equation
$$
\dfrac{G_n}{1-c}=\int\limits_{0}^{\infty}\Phi(\eta',\mu)a(\eta')d\eta'.
\eqno{(2.8)}
$$

Here $G_n=g_nl$, $g_n=\dfrac{d \ln n(y)}{dy}$, $l=v_T\tau$ is
the mean free path of the gas molecules,
$\tau=1/(\nu_1+\nu_2)$,
$c=\dfrac{\nu_1}{\nu_1+\nu_2}$, $\nu_1$ and $\nu_2$ is the frequency
of collisions between molecules of the first and
the second gas component.

We multiply equation (2.8) at the expression
$e^{-\mu^2}\gamma(\eta)\Phi(\eta,\mu)$ and integrate the
resulting equation by $\mu$. The result is that
$$
\dfrac{G_n}{1-c}\int\limits_{0}^{\infty}e^{-\mu^2}\gamma(\mu)
\Phi(\eta,\mu)d\mu=
$$
$$=\int\limits_{0}^{\infty}a(\eta')d\eta'
\int\limits_{0}^{\infty}e^{-\mu^2}\gamma(\mu)\Phi(\eta,\mu)
\Phi(\eta',\mu)d\mu,
$$
or, using the above notation and theorem,
$$
\dfrac{G_n}{1-c}(1,\Phi_\eta)=\int\limits_{0}^{\infty}a(\eta')
N(\eta-\eta')\delta(\eta-\eta')d\eta'.
$$

Hence, by the theorem 1 we have
$$
a(\eta)=\dfrac{G_n}{1-c}\dfrac{\eta}{N(\eta)}=\dfrac{G_n}{1-c}\cdot
\dfrac{\eta}{e^{\eta^2}\gamma(\eta)\lambda_c^+(\eta)\lambda_c^-(\eta)}=
$$
$$
=\dfrac{G_n}{1-c}\cdot\dfrac{e^{-\eta^2}}{X^+(\eta)\lambda_c^-(\eta)},
$$
which coincides exactly with the result of \cite{14}.

We introduce another scalar product on the set of their
eigenfunctions with the integration of the spectral parameter and weight
$r(\eta)=1/N(\eta)$
$$
\langle f,g\rangle=\int\limits_{0}^{\infty}\dfrac{1}
{N(\eta)}f(\eta)g(\eta)d\eta.
$$

Similarly can we prove

{\sc Theorem 3.} Eigenfunction of the characteristic equation
corresponding to the continuous spectrum, orthogonal and have
the relation
$$
\langle \Phi_\eta(\mu)\Phi_\eta(\mu')\rangle=\dfrac{1}{\rho(\mu)}
\delta(\mu-\mu').
\eqno{(2.10)}
$$

We represent in explicit form the equality (2.10):
$$
\langle \Phi_\eta(\mu)\Phi_\eta(\mu')\rangle \equiv
\int\limits_{0}^{\infty}\dfrac{e^{-\eta^2}}
{\lambda_c^+(\eta)\lambda_c^-(\eta)\gamma(\eta)}
\Phi_\eta(\mu)\Phi_\eta(\mu')d\eta=
$$
$$
=\dfrac{e^{\mu^2}}{\gamma(\mu)}\delta(\mu-\mu').
$$

From theorems 2 and 3 shows that in the transition to orthogonality
spectral parameter swapped weight and normalization
integral.

\begin{center}
  \bf 3. Kinetic equation with collision frequency proportional
to absolute velocity of the molecules
\end{center}

We now consider the equation (1.6). Substitution (2.1)
reduces this equation to the characteristic
$$
(\eta-\mu)\Phi(\eta,\mu)=\dfrac{3}{4}\eta
\eqno{(3.1)}
$$
with single normalization
$$
n(\eta) \equiv \int\limits_{-1}^{1}(1-\mu'^2)\Phi(\eta,\mu')d\mu'
\equiv 1.
\eqno{(3.2)}
$$

From the equations (3.1) and (3.2) find the eigenfunctions of
the characteristic equation
$$
\Phi(\eta,\mu)=\dfrac{3}{4}\eta P\dfrac{1}{\eta-\mu}+
\dfrac{\lambda(\eta)}{1-\eta^2}\delta(\eta-\mu),
\eqno{(3.3)}
$$
where $\lambda(z)$ is the dispersion function,
$$
\lambda(z)=1+\dfrac{3}{4}z\int\limits_{-1}^{1}
\dfrac{1-\tau^2}{\tau-z}d\tau=\dfrac{3}{4}\int\limits_{-1}^{1}
\dfrac{\tau(1-\tau^2)}{\tau-z}d\tau=
$$
$$
=-\dfrac{1}{2}+\dfrac{3}{2}(1-z^2)
\lambda_0(z),
$$
$$
\lambda_0(z)=1+\dfrac{z}{2}\int\limits_{-1}^{1}
\dfrac{d\tau}{\tau-z}=1+\dfrac{z}{2}\ln\dfrac{1-z}{1+z}.
$$

Discrete spectrum of the characteristic equation, as shown in
\cite{10,14}, consists of one point $\eta_i=\infty$ multiplicity two.
This point corresponds to the eigenfunction $\Phi_\infty=1$,
corresponding normalization $n(\eta)=\dfrac{4}{3}$.

Homogeneous Riemann boundary value problem
$$
\dfrac{X^+(\mu)}{X^-(\mu)}=\dfrac{\lambda^+(\mu)}{\lambda^-(\mu)},\qquad
0<\mu<1,
$$
as shown in \cite{10,14}, has a solution
$$
X(z)=\dfrac{1}{z}e^{V(z)},
\eqno{(3.4)}
$$
where
$$
V(z)=\dfrac{1}{\pi}\int\limits_{0}^{1}\dfrac{\theta(\mu)-\pi}{\mu-z}.
$$

Here
$\theta(\mu)=\arg \lambda^+(\mu)$, or
$$
\theta(\mu)=\arcctg\dfrac{4\lambda(\mu)}{3\pi \mu(1-\mu^2)}.
$$

Introduce the scalar product
$$
(f,g)=\int\limits_{0}^{1}(1-\mu^2)\gamma(\mu)f(\mu)g(\mu)d\mu,
$$
wherein
$$
\gamma(\mu)=\mu\dfrac{X^+(\mu)}{\lambda^+(\mu)}.
$$

Exactly the same way as theorem 1, we prove the following theorem.

{\sc Theorem 4.} The following relations hold
$$
(\Phi_\infty,\Phi_\infty)=-\dfrac{4}{3},
\eqno{(3.5)}
$$
$$
(\mu,\Phi_\infty)=-\dfrac{4}{3}V_1,
\eqno{(3.6)}
$$
where
$$
V_1=-\dfrac{1}{\pi}\int\limits_{0}^{1}[\theta(\mu)-\pi]d\mu\approx
0.581946\cdots,
$$
и
$$
(\mu,\Phi_\eta)=\eta, \qquad \eta>0.
\eqno{(3.7)}
$$

{\sc Proof.} We prove the equalities (3.5) and (3.6).
We use the integral representation \cite{14}
$$
X(z)=\dfrac{3}{4}\int\limits_{0}^{1}\dfrac{(1-\tau^2)\gamma(\tau)}
{\tau-z}d\tau.
\eqno{(3.8)}
$$

We expand the function $X(z)$ in a neighborhood of infinity.
We use the equations (3.8) and (3.4). As result we have
$$
X(z)=-\dfrac{1}{z}\cdot
\dfrac{3}{4}\int\limits_{0}^{1}(1-\mu^2)\gamma(\mu)d\mu-
$$$$-\dfrac{1}{z^2}\cdot
\dfrac{3}{4}\int\limits_{0}^{1}\mu(1-\mu^2)\gamma(\mu)d\mu-\cdots,
\qquad z\to \infty,
\eqno{(3.9)}
$$
and
$$
X(z)=\dfrac{1}{z}+\dfrac{V_1}{z^2}+\cdots,\qquad z\to \infty.
\eqno{(3.10)}
$$
From a comparison of the coefficients of series (3.9) and (3.10)
implies the equalities
$$
\dfrac{3}{4}\int\limits_{0}^{1}(1-\mu^2)\gamma(\mu)d\mu=-1
$$
and
$$
\dfrac{3}{4}\int\limits_{0}^{1}\mu(1-\mu^2)\gamma(\mu)d\mu=-V_1,
$$
which proves the equalities (3.5) and (3.6).

The other equalities are proved similarly to theorem 1.

{\sc Theorem 1.} Eigenfunctions of the continuous spectrum
are orthogonal to each other and have the following
orthogonality relations
$$
(\Phi_\infty,\Phi_\eta)=0,
\eqno{(3.11)}
$$
$$
(\Phi_\eta, \Phi_{\eta'})=N(\eta)\delta(\eta-\eta'),
\eqno{(3.12)}
$$
where
$$
N(\eta)=\gamma(\eta)\dfrac{\lambda^+(\eta)\lambda^-(\eta)}{1-\eta^2}.
$$

Theorem 5 is proved similarly to theorem 2.

We apply the developed theory to the solution of the
Kramers problem. In \cite{10,14} shown that the solution of
the Kramers problem reduces to the solution of the integral equation
$$
2U_0-2G_v\mu+\int\limits_{0}^{\infty}\Phi(\eta',\mu)a(\eta')d\eta'.
\eqno{(3.13)}
$$

Where $U_0$ is the unknown dimensionless sliding speed, and
$G_v$ is the specified far from the wall dimensionless mass
velocity gradient.

To find the sliding speed multiply equation (3.13)  to
$\rho(\mu)=(1-\mu^2)\gamma(\mu)$ and integrate by $\mu$ from
$0$ to $1$. As a result, we obtain the equation
$$
2U_0(1,1)-2G_v(1,\mu)+
\int\limits_{0}^{\infty}a(\eta')(1,\Phi_{\eta'})d\eta'=0.
\eqno{(3.14)}
$$
According to theorem 5 $(1,\Phi_{\eta'})=0$. Therefore from
the equation (3.14) in view of theorem 5 we derive known
of \cite{10,14} result
$$
U_0=\dfrac{(1,\mu)}{(1,1)}G_v=V_1 G_v.
$$

To find the coefficient of the continuous spectrum $a(\eta)$
multiply (3.14) by
$(1-\mu^2)\gamma(\mu)\Phi(\eta',\mu)$ and integrate by $\mu$ from
$0$ to $1$. As a result, we obtain the equation
$$
2U_0(1,\Phi_{\eta'})-2G_v(\mu, \Phi_{\eta'})+
\int\limits_{0}^{\infty}a(\eta')(\Phi_\eta,\Phi_{\eta'})d\eta'=0.
\eqno{(3.15)}
$$
According to theorem 5
$$
(1,\Phi_{\eta'})=0, \qquad (\mu,\Phi_{\eta'})=\eta,\qquad
(\Phi_\eta,\Phi_{\eta'})=N(\eta)\delta(\eta-\eta').
$$

Therefore from the equation (3.15) that
$$
a(\eta)=\dfrac{\eta}{N(\eta)}(2G_v)=
\dfrac{1-\eta^2}{X^+(\eta)\lambda^-(\eta)}(2G_v),
$$
that exactly coincides with the known result of \cite{10,14,15}.

\begin{figure}[h]
\begin{center}
\end{center}
\begin{center}
\end{center}
\end{figure}

\begin{center}
  \bf 4. Conclusion
\end{center}

In this paper we develop a theory of of the orthogonality
eigenfunc\-ti\-ons of the characteristic equations corresponding
to two kinetic equations. This theory is developed on the
positive real axis (and in the range of $0<\eta<1$) using
the Riemann boundary value problem \cite{Gakhov} with a
coefficient equal to the ratio of the boundary values of
the dispersion function on the cut. Orthogonality applied to
solving boundary value problems for the equations considered.

\end{document}